\documentclass[preprint]{elsart}
\usepackage{epsfig}

\usepackage{amssymb}
\usepackage{subfigure}

\begin{document}

\begin{frontmatter}



\title{Rectification effects in coherent transport through single molecules}


\author{Florian Pump and Gianaurelio Cuniberti}

\address{Institute for Theoretical Physics, University of Regensburg, D-93040 Regensburg}

\begin{abstract}
A minimal model for coherent transport through a donor/acceptor
molecular junction is presented. The two donor and acceptor sites
are described by single levels energetically separated by an
intramolecular tunnel barrier. In the limit of strong coupling to
the electrodes a current rectification for different bias voltage
polarities occurs. Contacts with recent experiments of molecular
rectification are also given.
\end{abstract}

\end{frontmatter}

\section{Introduction}\label{sect:Introduction}
In recent years, molecular electronics has gained considerable
attention due to the viability to measure quantum transport
observables through single molecules. In fact, various effects
typical of the physics of mesoscopic electron systems ranging from
Coulomb blockade effects with weak coupling between the molecule and
the leads to coherent transport through strongly bonded molecules
have been observed~\cite{nitzan03,cfr05}. However, it is desirable
to obtain clear relationships between truly intrinsic molecular
properties and measured effects. For example, vibron assisted
tunneling~\cite{WangLR05} or current switching due to a controllably
induced conformational change of the molecular
shape~\cite{Moresco01,Nazin04} have been observed. Another idea,
firstly proposed theoretically by Aviram and Ratner~\cite{AR74a}, is
to create molecules able to rectify electric currents, i.e. $I(U)
\neq -I(-U)$. This can be either achieved by distinct asymmetric
molecules (e.g. asymmetric tunneling barriers between the contacts
and a central part of the molecule~\cite{Kornil02}) or by an
intrinsic modification of the electronic properties of the
molecules~\cite{Metzger02,Elbing05}.

\section{System and Method}\label{sect:MethSys}

Having in mind the experiments by Elbing \textit{et
al}.~\cite{Elbing05}, where a molecule consisting of two broken
$\pi$-conjugated donor/acceptor subunits connected to gold
electrodes via thiol groups showed a diode-like behavior, a minimal
model for quantum transport through such a molecular system is
presented. Though lacking \textit{ab initio} accuracy, our model
gives a basic insight to the mechanism of molecular rectification. \\
We describe the donor-acceptor molecule by two single levels
separated by a tunneling junction. The on-site energies are
$\epsilon_{11}^0$ and $\epsilon_{22}^0$ (in the case of the isolated
molecule), respectively (see Fig. 1). These energies are measured
relatively to the Fermi energies of the electrodes, which are both
set arbitrarily to zero. The two parts of the molecule are coupled
by a weakly conducting bridge, which can be formed by a
$\sigma$-bond \cite{AR74a} or an almost broken biphenylic
$\pi$-bond~\cite{Elbing05}.
\begin{figure}[b]
\begin{center}
\subfigure[$U<0$]{\centering \psfig {file=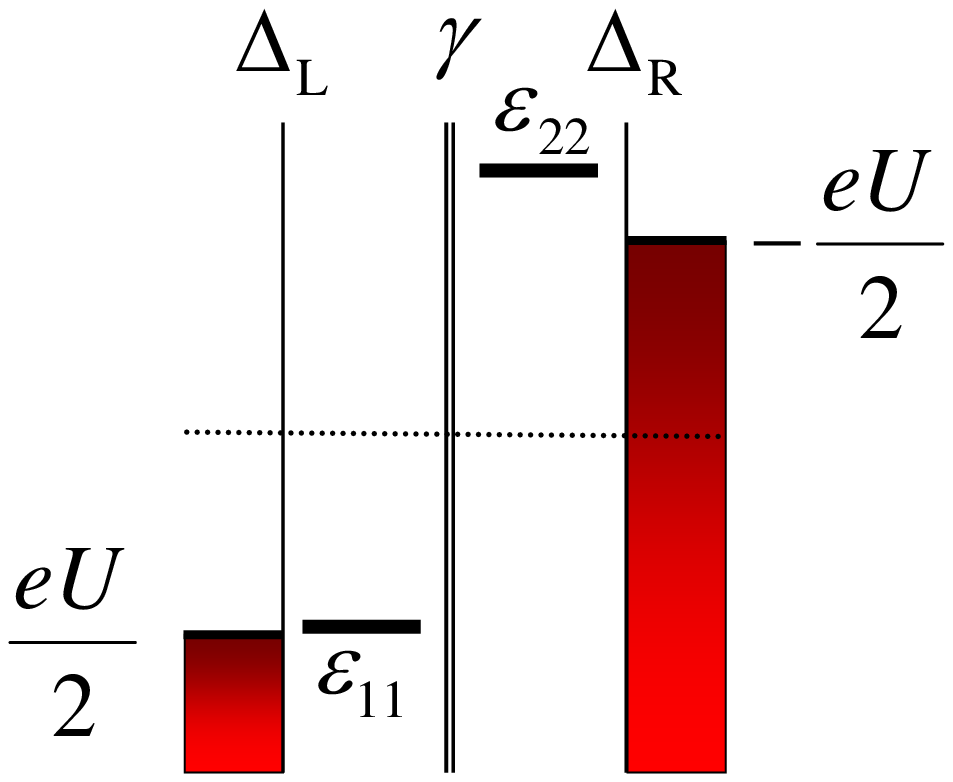,
height=4.1 cm} \label{fig:levelssz}}\subfigure[$U=0$]{\centering
\psfig {file=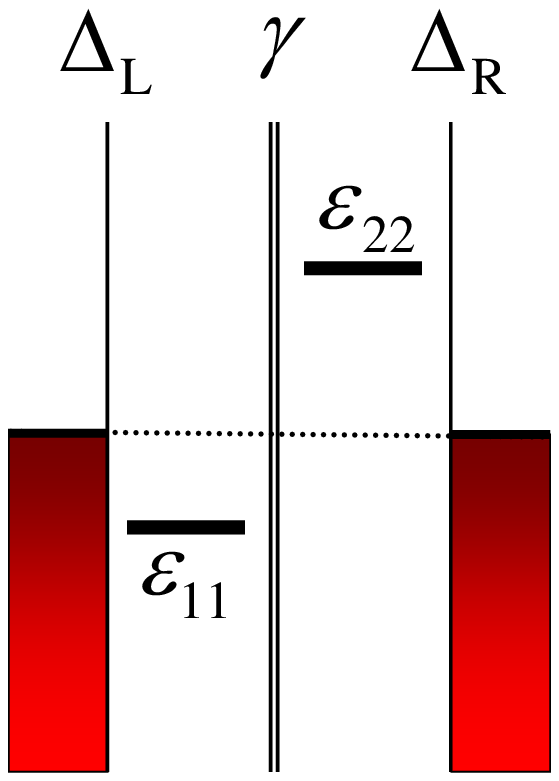,height=4.1 cm} \label{fig:levelsez}}
\subfigure[$U>0$]{\centering\psfig{file=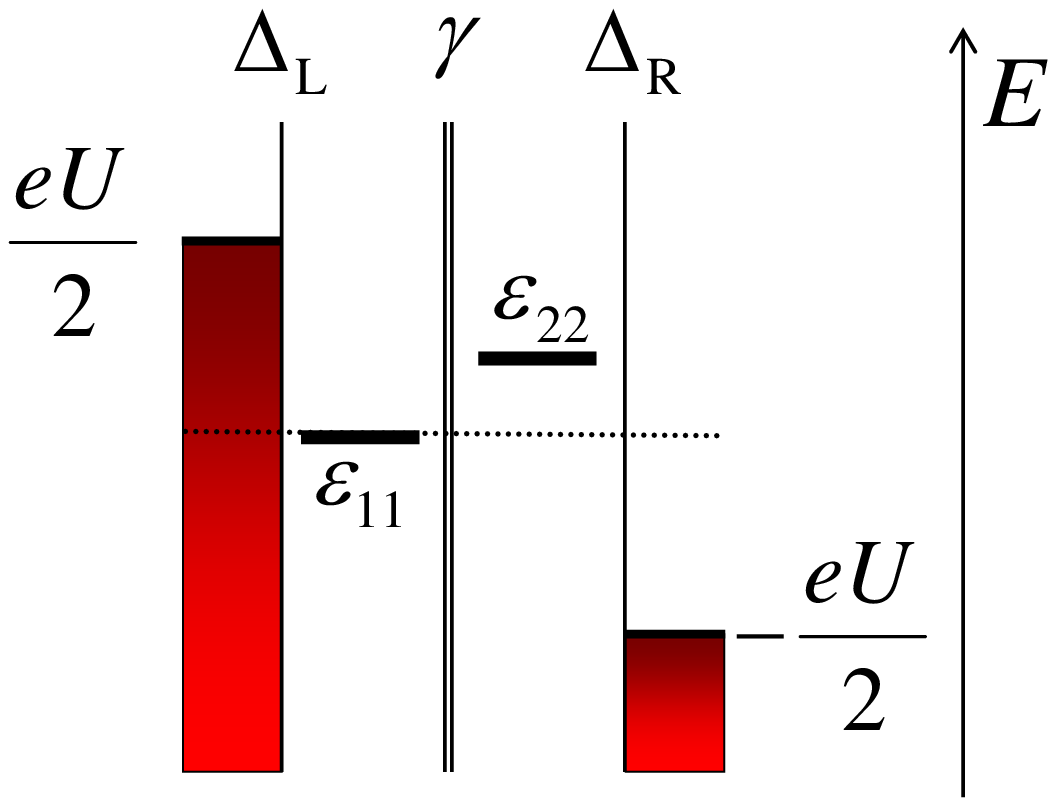,height=4.1
cm} \label{fig:levelslt}} \caption{Sketch of the energy levels at
negative (a), zero (b) and positive (c) bias voltage. The dotted
lines indicate the position of the Fermi energy. Since the molecular
levels themselves are affected by the bias voltage, Eq.
(\ref{eq:level_flow}), the distance and positions of the levels are
strongly sensitive to the bias polarity.}
\end{center}
\label{fig:levels}
\end{figure}
Under the external bias voltage $U$, which is applied symmetrically
on the two contacts, the chemical potentials of left and right lead
are moved: $\mu_{\mathrm{L}}=-\mu_{\mathrm{R}}=eU/2$. Furthermore,
the
 energy levels of the molecular sites are shifted as
\begin{equation} \label{eq:level_flow}
\epsilon _{11} \left( U \right)= \epsilon_{11}^{0}+\eta_1 eU , \quad
\epsilon _{22} \left( U \right)= \epsilon_{22}^{0}+\eta_2 eU.
\end{equation}
To implement the fact that the bottleneck tunneling barrier in the
system would tend to pin these levels to the nearby electrode, we
assume that one fourth of the applied voltage drops at the contacts
between the molecule and the leads and one half of the bias voltage
drops in the middle of the system between the two parts of the
molecule at the intra-molecular barrier, i.e. $\eta_1=-\eta_2=0.25$,
this assumption is then lifted at the end of this paper when
discussing different strengths of coupling between electrodes and
molecule.
\\The Hamiltonian describing the molecule is given by
\begin{equation}
H_{\mathrm{M}}=\sum_{\alpha,\beta=1,2}\epsilon_{\alpha \beta}\left(
U \right)
c_{\alpha}^{\dagger}c_{\beta}^{\phantom{\dagger}}+\textrm{H.c.},
\end{equation}
whereas the diagonal elements $\epsilon_{\alpha\alpha}$ are defined
by the on-site energies $\epsilon_{\alpha \alpha}\left( U \right)$
and the off diagonal elements $\epsilon_{12}$ and $\epsilon_{21}$
are given by the hopping parameter between the two levels,
$-\gamma$, which is a measure for the inter-site coupling. In the
two dimensional base of the localized orbital operators the
Hamiltonian matrix simply writes:
\begin{equation} \label{eq:HamilMatrix}
\bf{H}_{\mathrm{M}}= \left( \begin{array}{cc} \epsilon _{11} \left(
U \right) & -\gamma \\ -\gamma & \epsilon _{22} \left( U \right)
\end{array} \right).
\end{equation}
The total Hamiltonian contains two additional terms due to the
electrodes and their coupling to the molecule:
$H=H_{\mathrm{M}}+H_{\mathrm{M-leads}}+H_{\mathrm{leads}}.$
To calculate the spectral and transport properties of the molecule
coupled to leads, we use the nonequilibrium Green functions
technique for finite bias voltage without interaction.
The retarded Green function matrix of the molecular
region dressed by the electrode self-energies
reads~\cite{CunibertiGG02}
\begin{equation}
\left( \bf{G}^{\textrm{r}} \right)^{-1} =
\left(\left(E+\textrm{i}0^{+}\right)\bf{1}
-\bf{H}_{\mathrm{M}}-\bf{\Sigma}_{\mathrm{L}}-\bf{\Sigma}_{\mathrm{R}}\right),
\end{equation}
where
\begin{equation}
\bf{\Sigma}_{\mathrm{L}}=\left( \begin{array}{cc} -\textrm{i}
\Delta_\mathrm{L} & 0
\\0&0 \end{array} \right), \quad \bf{\Sigma}_{\mathrm{R}}=\left(
\begin{array}{cc} 0 & 0 \\0&-\textrm{i}\Delta_\mathrm{R} \end{array} \right)
\end{equation}
are the self-energy matrices of left and right lead, which already
lift the molecular resonances away from the real energy axis,
surrogating the need of a small imaginary shift applied in the
definition of the retarded Green function. For the self-energies we
use the wide band approximation, which assumes a purely imaginary
energy-independent self-energy. $\Delta_{\mathrm{L,R}}$ describe the
hopping between the contacts and the two energy levels. Both
$\Delta_{\mathrm{L}}$ and $\Delta_{\mathrm{R}}$ are assumed to be
constant positive real numbers with no energy dependence (wide band
approximation). This is a reasonable assumption when thinking of
gold electrodes whose bands are more extended than the molecular
active energetic window.
The transmission probability is obtained from the Green function by
the Fisher-Lee relation $T\left(E,U\right)= \left\{
\bf{\Gamma}_{\mathrm{L}} \bf{G}^{\mathrm{r}}
\bf{\Gamma}_{\mathrm{R}} \bf{G}^{\mathrm{r} \dagger}
\right\}$~\cite{FL81,Datta99} with the matrices
$\bf{\Gamma}_{\mathrm{L}}=\textrm{i} \left(
\bf{\Sigma}_{\mathrm{L}}^{\phantom{\dagger}}-\bf{\Sigma}_{\mathrm{L}}^{\dagger}
\right)$
and $\bf{\Gamma}_{\mathrm{R}}=\textrm{i} \left(
\bf{\Sigma}_{\mathrm{R}}^{\phantom{\dagger}}-\bf{\Sigma}_{\mathrm{R}}^{\dagger}
\right)$
as the anti-Hermitian parts of the self-energy matrices of the
contacts.
\\In the case of the two site-model studied here, it is easy to
obtain an analytical expression for the transmission probability as
a function of voltage and charge injection energy:
\begin{eqnarray}
T \left( E,U \right) &=& 4\Delta_{\mathrm{L}}\Delta_{\mathrm{R}}
\left| G_{12}^{\textrm{r}} \right| ^{2} = 4\Delta_{\mathrm{L}}
\Delta_{\mathrm{R}} \gamma^{2}/\left(A+B\right)
\\ A &=& \left[ \left( E-\epsilon_{11} \left( U \right) \right)\left(
E-\epsilon_{22} \left( U \right) \right)-\gamma^{2}-\Delta_{\mathrm{L}}\Delta_{\mathrm{R}}\right]^2 \nonumber \\
B &=& \left[\Delta_{\mathrm{L}} \left(E-\epsilon_{11} \left( U
\right) \right) +\Delta_{\mathrm{R}} \left(E-\epsilon_{22} \left( U
\right) \right) \right]^2. \nonumber
\end{eqnarray}
Knowing the transmission probability, the current through the system
can be obtained by the relation~\cite{Datta99}:
\begin{equation} \label{eq:current}
I\left(U \right)=\frac{2e}{h}\int \mathrm{d}E \;\;T \left(E,U
\right) \left(f_{\mathrm{L}} \left( E \right) -  f_{\mathrm{R}}
\left( E \right) \right).
\end{equation}
Here, $f_{\mathrm{L},\mathrm{R}} \left( E
\right)=f\left(E-\mu_{\mathrm{L},\mathrm{R}}\right)$ are the Fermi
functions of left and right electrode depending on the bias voltage
and the temperature, which in our calculations is set to $30$~K.
To understand the evolution of the molecular levels under the
influence of the applied bias voltage and the coupling to the leads,
we analyze the density of states (DOS), especially the localized
density of states (LDOS) projected on the two sites of the
molecule:\begin{eqnarray}
\mathrm{LDOS}_{\mathrm{1}}\left(E,U\right)=- \frac{1}{2
\pi}\textrm{Im}\, \textrm{Tr}_{1}G^{\textrm{r}}= -\frac{1}{2 \pi}
\textrm{Im}
G^{\textrm{r}}_{11}, \nonumber \\
\mathrm{LDOS}_{\mathrm{2}}\left(E,U\right)=- \frac{1}{2
\pi}\textrm{Im}\, \textrm{Tr}_{2}G^{\textrm{r}}= -\frac{1}{2 \pi}
\textrm{Im} G^{\textrm{r}}_{22}.
\end{eqnarray}
Additionally, we define $\Delta_{\mathrm{DOS}} =
\mathrm{LDOS}_{\mathrm{1}}-\mathrm{LDOS}_{\mathrm{2}}$ and
$\Pi_{\mathrm{DOS}}=\mathrm{LDOS}_{\mathrm{1}}\cdot\mathrm{LDOS}_{\mathrm{2}}$.
 Looking at $\Delta_{\mathrm{DOS}}$, one can find the localization of a molecular level: positive
$\Delta_{\mathrm{DOS}}$ implies that the state is more localized on
site $1$, negative $\Delta_{\mathrm{DOS}}$ points to a pronounced
state on site $2$. $\Pi_{\mathrm{DOS}}$, on the contrary, enhances
with the equipartition of molecular states on the two sites.
\section{Results and Discussion}\label{sect:Results}
\begin{figure}[t]
\centerline{\subfigure{\psfig{file=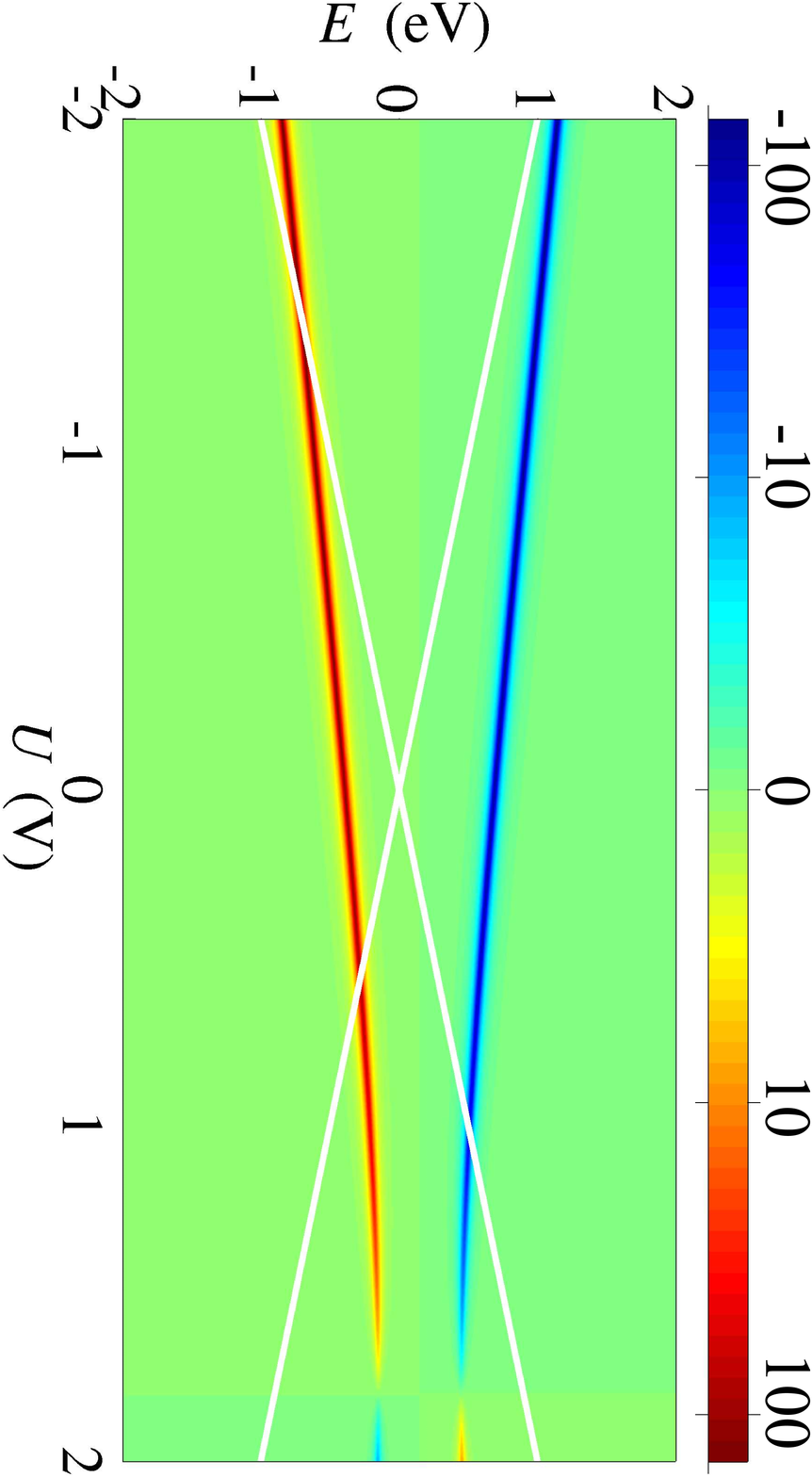,height=1.1
\textwidth, angle=90}}} \label{fig:DOS_diff}
\centerline{\subfigure{\psfig{file=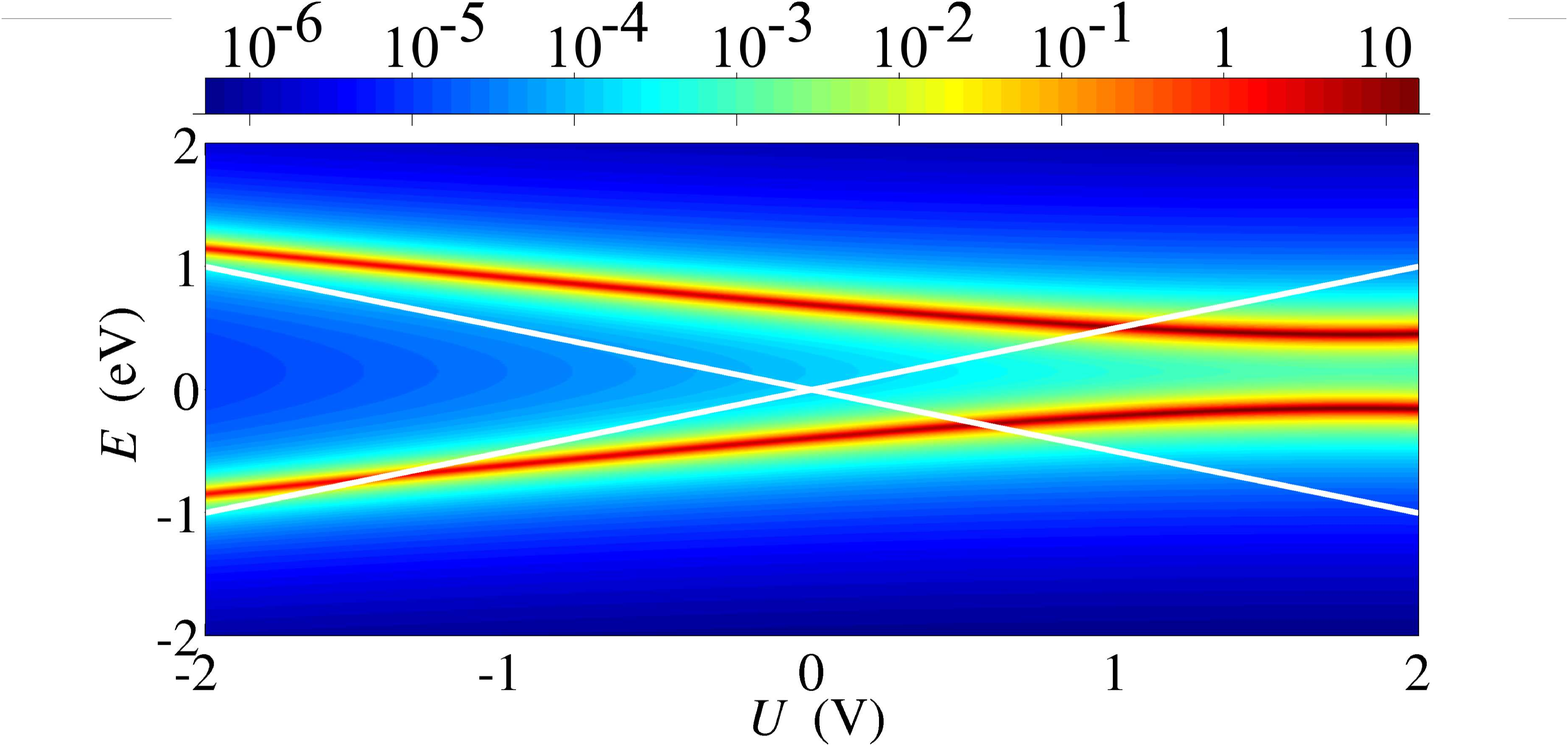,
width=1.1\textwidth, angle=0}}} \label{fig:DOS_prod}
\caption{Difference (upper panel) and product (lower panel) of the
localized density of states of site $1$ and site $2$ as a function
of the bias voltage and the energy. The white lines indicate the
bias window. In order to emphasize the molecular states, a
logarithmic nonlinear scale has been applied. (Parameters:
$\epsilon_{11}^{0}=-0.3$~eV, $\epsilon_{22}^{0}=0.6$~eV,
$\Delta_0=0.02$~eV, $\gamma=0.3$~eV, $\eta_1=-\eta_2=0.25$).}
\label{fig:DOS}
\end{figure}

The case of two identical sites
$\epsilon_{11}^{0}=\epsilon_{22}^{0}$, independently of their value,
leads straightforwardly to a symmetric behavior of the current with
respect to an inversion of the applied bias, i.e.
$I\left(U\right)=-I\left(-U\right)$. First, we want to study the
case of two levels with different energies coupled symmetrically to
the leads ($\Delta_{\mathrm{L}}=\Delta_{\mathrm{R}}=\Delta_0$).
In the upper panel of Fig. \ref{fig:DOS}, the difference of the DOS
projected on the two levels is shown as a function of the applied
bias voltage and the energy: the two lines show the variation of the
two molecular levels which are broadened by the coupling to the
leads. The localization of the levels is indicated as follows: red
resp. black color of the central part indicates that the level is
situated on site 1, blue points to a level localized on site 2. For
large negative bias voltages, the two levels are far apart. With
increasing voltage, the two levels approach each other, but do not
cross and move away from each other again. This repulsion is
quantified by the inter-site coupling $2\gamma$. At this voltage,
$\Delta_{\mathrm{DOS}}$ is zero whereas $\Pi_{\mathrm{DOS}}$ takes
its maximum (lower panel). After the avoided crossing, a change in
the localization of the energy levels takes place.
\begin{figure}[t]
\centerline{\psfig{file=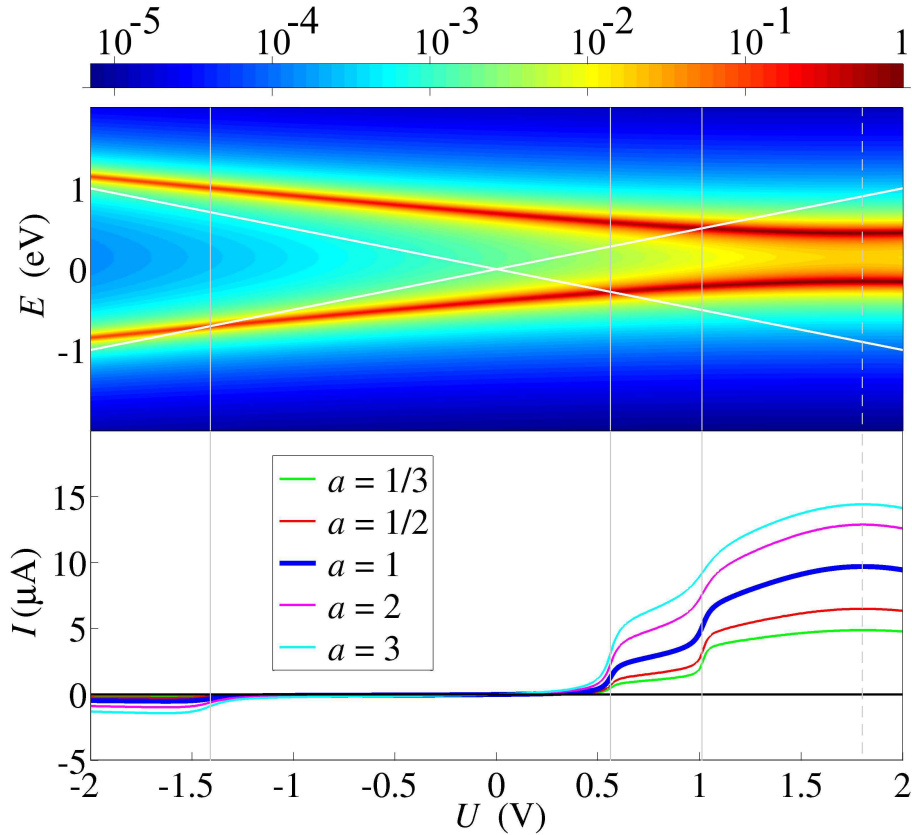, height=1\textwidth,
angle=270}}
 \caption{Transmission function $T\left(E,U\right)$ (upper
panel, in a logarithmic scale) and current $I\left(U\right)$ (lower
panel) for a two-site system. The white lines in the upper panel
indicate the bias window. Different rectification degrees are
obtained for different asymmetry parameters $a$, with
$\Delta_{\mathrm{L}}=\Delta_0$ and $\Delta_{\mathrm{R}}=a\Delta_0$.
(Parameters: $\epsilon_{11}^{0}=-0.3$~eV,
$\epsilon_{22}^{0}=0.6$~eV, $\Delta_0=0.02$~eV, $\gamma=0.3$~eV,
$\eta_1=-\eta_2=0.25$).}
 \label{fig:TransCurr}
\end{figure}

Fig.~\ref{fig:TransCurr} shows the calculated transmission function
and the current flowing through the system (thick blue line in the
lower panel) calculated using Eq.~(\ref{eq:current}). The bias
window, indicated in the plot of the transmission function by the
two white lines, is determined by the integration interval fixed
because of the Fermi functions at low temperature. A step in the
current-voltage curve appears, when a level enters the bias window,
as it happens once at negative and once at positive bias voltages.
If $\Delta_0$ is small compared to the level spacing, i.e. for
relatively sharp molecular levels, the positions of the steps is
given by the crossing of the eigenenergies (this can be determined
analytically by a diagonalization of the Hamiltonian in Eq.
(\ref{eq:HamilMatrix})) and the lines defining the bias window. In
our case, as depicted by the solid grey vertical lines in Fig.
\ref{fig:TransCurr}, this takes place at $U=-1.41$~V in negative
bias direction and at $U=0.56$~V and $U=1.01$~V for positive bias.
The height of the steps however depends crucially on the distance
between the two levels: in the region of negative voltage, the two
states are far apart and the height of the steps is therefore
relatively small, whereas for positive bias voltage, the levels are
much closer to each other, which results in a more resonant state
and a much higher current. This explains the observed rectification
in our model. The highest current is reached where the two levels
are closest to each other. The peak in $I(U)$ appears at the bias
voltage $\left(\epsilon_{22}^{0}-\epsilon_{11}^{0}\right) /
e\left(\eta_1-\eta_2\right)$ (in our case $U= 1.80$~V, depicted by
the dashed grey line in Fig. \ref{fig:TransCurr}) which can be
obtained by a minimization of the energy gap between the two
eigenenergies of the bare molecule~\cite{Lakshmi05}. For higher
voltages, the distance between the levels grows again and the
current decreases.
\newline
So far, we only considered symmetric coupling of the molecule to the
leads. However, in experimental investigations of electronic
transport through single molecules, the strength of the bonds
between the molecule and the leads is mostly not well defined.
Therefore, it is important to study the case of the molecule coupled
asymmetrically to the electrodes, i.e.
$\Delta_{\mathrm{L}}=a_{\mathrm{L}}\Delta_0$ and
$\Delta_{\mathrm{R}}=a_{\mathrm{R}}\Delta_0$. The thin lines in Fig.
\ref{fig:TransCurr} represent the values of the current for
asymmetric coupling $a_{\mathrm{L}}=1$ and different values of
$a_{\mathrm{R}}=a$, resulting in $\Delta_{\mathrm{L}}=\Delta_0$ and
$\Delta_{\mathrm{R}}=a\Delta_0$. The other model parameters remain
unchanged, especially $\eta_1=-\eta_2=0.25$. In this case, the
general shape of the current curve does not change; the steps and
the maximum in the current voltage curve do not change their
positions, only the absolute values of the current are different due
to the varying coupling to the leads. To understand the correct
voltage drop along the molecule, one should use the Poisson equation
with boundary conditions given by the applied bias voltage. This
goes beyond the scope of this paper which concentrates on a
simplified picture of the molecular electronic structure. Thus, in
order to improve the plain effect of asymmetric coupling to the
shape of the voltage drop along the system, we assume a larger
voltage drop at the weaker contact. This changes as a consequence
the values of the factors $\eta_{1,2}$ defined in
Eq.~(\ref{eq:level_flow}) according to, e.g,:
$\eta_{1}=\left(3a_{\mathrm{R}}-a_{\mathrm{L}} \right)/8$ and
$\eta_{2}=\left(a_{\mathrm{R}}-3a_{\mathrm{L}} \right)/8,$ which for
the case $a_{\mathrm{L}}=a_{\mathrm{R}}=1$ would restore the
previously used values of $\eta_{1,2}$.
\begin{figure}[t]
\centerline{\psfig{file=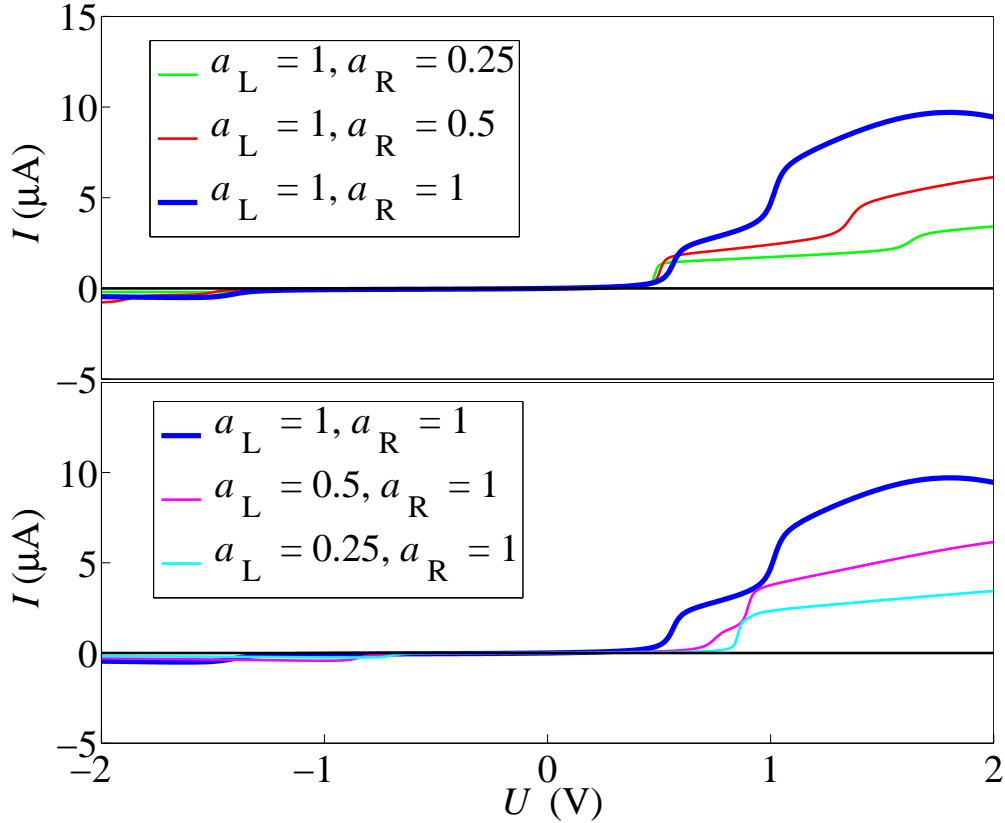, height=1.1\textwidth, angle=270}}
\caption{Current voltage characteristics for two cases of asymmetric
coupling. In the upper panel, $a_{\mathrm{L}}=1$, the lower panel
shows the case $a_{\mathrm{R}}=1$. (Parameters:
$\epsilon_{11}^{0}=-0.3$~eV, $\epsilon_{22}^{0}=0.6$~eV,
$\Delta_{0}=0.02$~eV, $\gamma=0.3$~eV, $\eta_1=-\eta_2=0.25$).}
\label{fig:avar}
\end{figure}
The results for this refined model are shown in Fig~\ref{fig:avar},
the thick blue line again representing the symmetric case. Here, the
position of the steps is shifted to different voltages. As it
becomes evident from a comparison of Figs. \ref{fig:TransCurr} and
\ref{fig:avar}, different shapes of the voltage drop imply a shift
of the position where the current significantly changes, i.e. where
the steps appear. Rectification effects due to the asymmetric
coupling as shown in Fig.~\ref{fig:TransCurr}, have been already
reported in the literature~\cite{DattaRoi97} but we here we want to
stress that the precise value of the current jumps are \textit{not}
a mere electronic structure effect, but also depend on the profile
of the electric field under the fixed applied bias voltage.

\section{Conclusions and Outlook}\label{sect:Conclusios}
In this paper, we have shown a minimal model for rectification
effects in coherent transport through single molecules which was
inspired by recently published experiments~\cite{Elbing05}. An
extension to this model includes charging effects~\cite{Song06},
taking into account on-site correlations on the two molecular sites,
which becomes important in the case of a weak coupling between the
molecule and the electrodes. These implementations though do not
essentially change the nature of the effects observed in the picture
discussed here, except for a natural splitting of conductance steps
due to the lifting of the spin degeneracy induced by the correlation
effects.
Additionally, density functional theory can be used to calculate the
positions of the energy levels of the molecule and their exact
behavior under each bias voltage applied to the molecule, combined
with a calculation of the correct voltage drop distribution at this
voltage solving the Poisson equation. Together with nonequilibrium
Green functions, this procedure allows to calculate the transmission
$T\left(E,U\right)$ and the current $I\left(U\right)$, and not only
the electronic structure under an applied voltage, in a truly
\textit{ab initio} way. Such study is currently under
investigation~\cite{Pump}.

\section{Acknowledgements}\label{sect:ack}
We thank Bo Song and Dmitry A. Ryndyk for useful discussions. This
work has been supported by the Volkswagen Foundation grant Nr.
I/78-340.



\bibliographystyle{elsart-num}

\end{document}